# A Wavelet Approach for the Estimation of Left Ventricular Early Filling Wave Propagation Velocity from Color-M-Mode Echocardiograms

**Brief Title (not more than 45 characters) = New Measure of Filling for LV Diastolic Dysfunction**

**Authors:**

Sreyashi Chakraborty[1], Hiroyuki Iwano[2,3], Michael E. Hall, MD[2], Pavlos P. Vlachos[1][$]

**In memoriam, William Campbell Little, M.D. passed away July 9, 2015.**

**Affiliations:**

[1]School of Mechanical Engineering, Purdue University

[2] Division of Cardiology, Department of Medicine, University of Mississippi Medical Center.

[3] Department of Cardiovascular Medicine, Faculty of Medicine and Graduate School of Medicine, Hokkaido University, Sapporo, Japan

[$]Correspondence to: Pavlos Vlachos, School of Mechanical Engineering, Purdue University, pvlachos@purdue.edu

Word Count = 5244

## Abstract

**Objective:** This work evaluates a new approach for calculating the left-ventricular (LV) early filling propagation velocity ($V_P$) from color M-Mode (CMM) echocardiograms using wavelet analysis. Unlike current approaches, the method requires no assumptions, user inputs, or heuristic conventions.

**Background:** Current methods for measuring $V_P$ using CMM echocardiography do not account for the spatiotemporal variation of the filling wave propagation velocity. They are instead confined by empirical assumptions and user inputs that significantly hinder the accuracy of $V_P$, subsequently limiting its clinical utility.

**Methods:** We evaluated three methods for measuring LV early filling $V_P$: conventional $V_P$, the strength of propagation ($V_S$), and $V_P$ determined from the most energetically important wave (Peak-$V_W$), using 125 patients (Group A) with normal filling (n=50), impaired relaxation (n=25), pseudonormal filling (n=25) and restrictive filling (n=25), and in 69 patients (Group B) with normal (n=32), dilated (n=15), and hypertrophic ventricles (n=22).

**Results:** Peak-$V_W$ most accurately distinguished normal left ventricles from diseased ones. Specifically, for Group A which spanned the stages of diastolic function, using receiver operator characteristics (ROC) and measuring their corresponding area under the curve (AUC), the AUC for Peak-$V_W$ was 0.92, versus 0.62 for conventional $V_P$, 0.63 for $V_S$ and 0.58 for intraventricular pressure difference (IVPD). These correspond to a 50-70% improvement in classification ability. Similar improvements were found using Group B.

**Conclusion:** A new determination of the LV early filling using wavelet analysis, Peak-$V_W$, may provide a more accurate evaluation of diastolic function than the standard method of determining $V_p$ and enable better diagnostic classification of patients suffering with diastolic dysfunction.

# Condensed Abstract

A new metric (Peak-$V_W$) for calculating the left-ventricular (LV) early filling propagation velocity from color M-Mode (CMM) echocardiograms using wavelet analysis is proposed. It accurately distinguished normal left ventricles from diseased ones. Specifically, for a group which spanned the four stages of diastolic function, using receiver operator characteristics (ROC) and measuring their corresponding area under the curve (AUC), the AUC for Peak-$V_W$ was 0.92, versus 0.62 for conventional propagation velocity $V_P$, 0.63 for filling strength $V_S$ and 0.58 for intraventricular pressure difference (IVPD). These correspond to a 50-70% improvement in classification ability of patients suffering from diastolic dysfunction.

# Abbreviations list

| | |
|---|---|
| LV | left ventricle |
| CMM | color M-mode |
| Peak-$V_W$ | wavelet based early filling propagation velocity |
| $V_P$ | early filling propagation velocity |
| $V_S$ | filling strength |
| IVPD | intraventricular pressure difference |
| LVMI | left ventricle mass index |
| DCM | dilated cardiomyopathy |
| HCM | hypertrophic cardiomyopathy |
| LVH | left ventricle hypertrophy |
| N | normal |
| IR | impaired relaxation |
| P | pseudonormal |
| R | restrictive |
| SI | sphericity index |
| EF | ejection fraction |
| CWT | continuous wavelet transform |
| IQR | inter-quartile range |
| AUC | area under curve |
| ROC | receiver-operator characteristic |
| n | number of patients |

# Introduction

Diastolic function of the left ventricle (LV) can be evaluated from color M-mode (CMM) Doppler echocardiography, which records a spatiotemporal map of blood velocity along a scan-line from the mitral valve towards the apex [1]. One or more combination of hydrodynamic parameters that characterize ventricle filling are conventionally calculated from the CMM scans. Some standard metrics from CMM scans used to assess LVDD are the propagation velocity ($V_P$) and the intraventricular pressure difference (IVPD) [2-5]. In chronic heart failure (HF) patients regardless of LV ejection fraction (EF), early diastolic IVPD during exercise was found to be closely related to the exercise capacity [6]. Apical IVPD (mid-LV to apex) was found to reduce total IVPD in HF patients while the basilar IVPD (left atrium to mid LV) was maintained by elevated LA pressure in HF patients [7]. Previous work from our group suggested improvements on the IVPD measurement by calculating it from a representative beat reconstructed by combining multiple heartbeats of the same patient and rectifying low resolution of CMM scans[8]. MRI based IVPD calculations were shown to be more accurate than CMM IVPDs when both were compared against computational fluid dynamic (CFD) measurements of IVPD as the ground truth [9]. IVPD calculations from 2D phase contrast magnetic resonance imaging scans showed a correlation with vortex ring formation at the mitral valve tips and its subsequent propagation into the ventricle [10]. The $V_P$ of early diastolic flow from the mitral valve toward the LV apex is conventionally measured as the slope of an iso-velocity contour of the blood velocity map, typically set at 50% of the maximum inflow velocity. This is performed using the slope of an iso-velocity contour determined by the CMM aliasing boundary (typically shown as a yellow to blue transition on the CMM colormap) from the mitral plane to a distance of 4 cm into the LV [11]. This is based on the assumption that the filling wave has a constant $V_P$ throughout early diastole. However, the iso-velocity contour is curvilinear, indicating that the propagation velocity is not constant during early filling [12]. In previous work, we found that describing $V_P$ using two slopes, one corresponding to a rapid initial and one a slower terminal $V_P$, provided a more accurate representation of early filling [13]. Further, using $V_S$, the product of the initial $V_P$ and the distance it travels into the LV provides a small improvement in assessing diastolic function [13]. Although this simple correction improves $V_P$ estimation, it does not capture the physical process by which the $V_P$ varies smoothly during the early filling process. Moreover the $V_P$ depends on the choice of aliasing boundary, corresponding to different iso-velocity contours [14,15], indicating that early diastolic LV filling is not a bulk wave moving with a constant velocity but instead is comprised of a range of spatiotemporally varying propagation velocities.

The objective of this work is to develop a new approach for the calculation of the LV early filling propagation velocity from CMM echocardiograms free of assumptions and heuristics, that is able to capture the underlying physics with high fidelity. The proposed method uses wavelets to analyze the early diastolic LV filling wave from CMM scans, and does not use any of the assumptions inherent in the traditional calculation of $V_P$, like subjective selection of iso-velocity contours, or measuring a slope based on an iso-velocity contour. In this work, we demonstrate that this automatic, objective, reproducible and user independent approach yields a more accurate, robust, and physically consistent estimation of $V_P$.

## Methods

### CMM Acquisition

The study was conducted according to protocols approved by the Virginia Tech and Wake Forest

University Baptist Medical Center Institutional Review Boards (IRB). Patients were all undergoing clinically indicated echocardiography at the Wake Forest University Baptist Medical Center and the data were de-identified prior to processing.

Doppler echocardiographic examinations were completed using an iE33 (Philips Medical Systems, Andover, Massachusetts) and a multiple frequency transducer. CMM echocardiograms were recorded in the apical four-chamber view with a color scale that optimized visualization as judged by the recording sonographers. The velocity map in all CMM scans were reconstructed using a dealiasing technique [13,16] before calculating the flow metrics like $V_P$, $V_S$, $V_W$ and IVPD .

## Patient Cohorts

Two patient groups were used in this analysis: Group A, for comparison of different diastolic function categories (n=125) and Group B, for comparison of different LV geometries (n=69). To evaluate LV diastolic dysfunction, Group A consisted of the following categories:

- Normal (N) filling, n=50

- Impaired relaxation (IR), n=25

- Pseudonormal (P) filling, n=25

- Restrictive (R) filling, n=25.

Patient characteristics for each group are provided in Table 1. Patients were classified based on clinically diagnosed diastolic function stage according to the guidelines [1].

To evaluate heart failure with structural remodeling, Group B is comprised of three patient cohorts (Table 2) that are classified on the basis of left ventricular geometry:

- Normal (N) cohort (n=32), without any abnormal echocardiographic findings including a normal filling pattern based on peak early filling mitral inflow velocity and normal mitral annular tissue Doppler velocities, [1].

- Severe LV hypertrophy (LVH) cohort (n=22) indicated by an elevated LV mass index (LVMI) with a cutoff value of $115/m^2$ for male and $95g/m^2$ for female [17].

- Dilated cardiomyopathy (DCM) cohort (n=15) based on a reduced sphericity index (SI) (SI<1.6) [18] and a reduced ejection fraction (EF) (<40%) and based on the clinical diagnosis by the physicians.

## Continuous Wavelet Transform Applied on Wave Propagation Velocity

Wavelet decomposition analysis was used to measure the $V_P$ of each wave component present in the filling wave. Fourier methods are typically used to identify what wavenumbers are present in a signal. However, the Fourier transform does not provide space or time localization for each wave component, which is needed here. The continuous wavelet transform (CWT), provides spatial and temporal localization of all wavenumbers, which enables us to calculate a new $V_P$ independent of iso-velocity assumptions.

The CWT, given in Equation 1, convolves a signal $f(x)$ with a series of scaled and translated continuous mother wavelet function, $\psi(\frac{x-b}{a})$, centered at $x$, to calculate the signal energy as a function of space and wavenumber. The scaling parameter, $a$, stretches or compresses the mother wavelet to capture different wavenumbers and translates it by $b$ to provide space localization.

$$W_T(a,b) = a^{-1/2} \int f(x) * \psi(\frac{x-b}{a})\, dx \qquad\qquad 1$$

A stretched wavelet will capture the energy of low wavenumber components, while a compressed wavelet captures the energy of high wavenumber components. The translation parameter of this function, $b$, shifts each wavelet along the entire length of the signal.

The results of a CWT are dependent on the choice mother wavelet and scales. The complex Morlet wavelet, Equation 2, which was chosen for this analysis, is a sine wave modulated by a Gaussian window. Its utility in biological applications is enhanced by its smooth continuous oscillations, as has been previously demonstrated [19].

$$\Psi(x) = \frac{1}{\sqrt{\pi f_b}} e^{i2\pi f_o x} e^{\frac{-x^2}{f_b}} \qquad\qquad 2$$

The frequency parameter, $f_o$, defines the frequency of the sine wave component and the bandwidth parameter, $f_b$, defines the bandwidth of the Gaussian modulating window. By controlling these two parameters, the mother wavelet can be optimized for the given input, with the stipulation that the mother wavelet must have near zero mean and have endpoints with zero magnitude [20,21].

The relationship of the input parameters can be expressed by:

$$f_b = \frac{n_{cycles}}{2\pi f_o} \qquad\qquad 3$$

We optimized the frequency parameter and then solved for a bandwidth parameter to maintain six (6) oscillations in the wavelet function.

The minimum wavenumber of interest is set to the minimum resolvable wavenumber from the CMM echocardiogram, or the inverse of the physical length of the CMM image. The maximum wavenumber to scale the mother wavelet is set to the wavenumber that represents an estimated signal-to-noise ratio cut-off. One hundred equally spaced scaling factors were generated to scale the mother wavelet from the minimum wavenumber to the maximum wavenumber.

## Calculating Wave Component Propagation Velocity

The result of the CWT applied to a spatiotemporal map of velocity values is a three-dimensional CWT power spectrum, localized in space (x), time (t), and wavenumber (k) as shown in Figure 1. The phase lag of each wavenumber between two time steps is performed using a phase only cross correlation [22]. The phase lag was measured by locating the maximum on the cross correlation. This phase lag, $\Delta x$ and the known time difference between the two time steps, ($\Delta t$), was used to calculate a propagation velocity corresponding to each wavenumber (k), $V_{w,k} = \Delta x/2\pi k \Delta t$. The time step ($\Delta t$) was manually chosen to be different for different disease stages.

As a byproduct of the CWT, all orthogonal waves whose linear superposition comprises the entire filling wave are obtained. This allows for any wave component of physical interest or any combination of those waves to be directly extracted, and the wave-specific $V_P$ be estimated. This capability provides the means to eliminate heuristic selections of filling velocity iso-contours and calculate $V_P$ for each wavenumber. We used the $V_P$ of the most energetically dominant wavenumber (Peak-$V_W$) to characterize early diastolic filling (Figure 2). For the remainder of this paper, we refer to this parameter as 'Peak-$V_W$'.

## Conventional Propagation Velocity ($V_p$) and Filling Strength ($V_s$)

The conventional $V_P$ parameter [23] was calculated as previously described in [13]. The filling strength parameter, ($V_S$) was calculated as the product of the initial $V_P$ and the distance it extended into the LV [13]. Note that both of these calculations are performed using automated computer programs increasing objectivity and reducing variability [13].

## CMM Derived IVPD

A Doppler-derived IVPD was computed using the CMM echocardiogram velocity data and the Euler equation [4,24], given in Equation 4.

$$\frac{\partial P}{\partial x} = -\rho \left[ \frac{\partial U}{\partial t} + U \frac{\partial U}{\partial x} \right] \qquad\qquad 4$$

where $U$ is velocity, $P$ is pressure and $\rho$ is the density of blood. Spatial and temporal derivatives are calculated and used to solve for $\frac{\partial P}{\partial x}$. The line integral along the length of the ventricle is calculated at each time step, yielding a spatio-temporal profile of pressure values $P(x, t)$. A temporal profile of IVPD is calculated by subtracting the pressure at the mitral valve from the pressure at the apex. The peak IVPD during early filling is then identified. This method has been validated with direct invasive pressure measurements by micro manometers [3,16,24].

## Statistical analysis

All patient cohort data are expressed as mean ± standard deviation. Medians and interquartile ranges (IQR) were also analyzed. We analyzed differences among groups using the Tukey-Kramer Honestly Significant Difference (HSD) test, which compared means of each pair of patient subgroups. The ability of the metrics to distinguish between patient sub-groups was analyzed using receiver-operator characteristic curves (ROC), specifically by comparing the area under the ROC curves (AUC). All data analyses were performed using Matlab (Mathworks).

# Results

## Intra-Operator and Inter-Operator Variability Analysis

Three different observers analyzed a random set of 15 beats for the inter-operator variability analysis and one observer analyzed 3 beats 15 times for intra-operator variability analysis. Table 3 displays the results of this analysis for each observer. The median coefficients of variation are calculated according to the following equation and reported as a variability metric:

$$Coefficient\ of\ Variation\ (\%) = \frac{100}{N} \sum_{i=1}^{N} \frac{standard\ deviation}{median}$$

Peak-$V_W$, had the smallest variability of 3.67 % intra-operator and 10.1% inter-operator, a 60-200% improvement compared to the conventional parameters $V_P$, $V_S$ and IVPD. Moreover, it should be noted that the $V_P$, $V_S$ and IVPD are also calculated with our automated software and have higher repeatability than traditional manual calculations.

## Group A: Diastolic Dysfunction

Peak IVPD, conventional $V_P$, filling strength $V_S$ and Peak-$V_W$ values for the cohort are plotted in Figure 3. Table 4 shows the medians and the IQRs for each group. IVPD was reduced in the IR group compared to the N group (1.81 +/- 1.5 mmHg versus 2.66 +/- 1.97 mmHg, respectively). There was a progressive increase in IVPD as diastolic dysfunction stage worsened from IR to P to R (Table 4). $V_P$, $V_S$ and Peak-$V_W$ were all lower in groups with diastolic dysfunction compared to normal patients. However, as shown from both Figure 3 and Table 4 only Peak-$V_W$ showed a consistent decrease with worsening diastolic dysfunction stage.

Table 5 quantifies the ability of each method to distinguish between diastolic function groups. Peak-$V_W$ was able to differentiate all of the disease subgroups (6 out of 6 pairs) ($p<0.0001$ for N vs P, N vs IR and N vs R, $p=0.006$ for P vs IR, $p<0.0001$ for IR vs R and $p=0.0015$ for R vs P). Conventional $V_P$ did not differentiate any pair of disease stages. IVPD differentiated 2 pairs ($p=0.0024^*$ for N vs IR and $p=0.0185^*$ for IR vs R) while $V_S$ differentiated 1 out of 6 pairs of disease stage combinations ($p=0.0146$ for P vs R).

The AUC on the ROC curves are plotted in Figure 4. In this plot the diseased state includes all patients with diastolic dysfunction of IR, P, and R. The Peak-$V_W$ (AUC=0.92) showed significantly improved classification ability over the Doppler-derived IVPD (AUC=0.58), the conventional parameter $V_P$ (AUC=0.62), and $V_S$ (AUC=0.63).

## Group B: Dependence on Left Ventricle Geometry

Peak IVPD, conventional $V_P$, filling strength $V_S$ and wavelet based Peak-$V_W$ values for patient cohorts corresponding to Normal, DCM and LVH groups are shown in Figure 5. The group median and the inter-quartile ranges are reported in Table 6. Both diseased patient cohorts displaying LV remodeling (DCM and LVH) showed decreased values for all methods compared to the Normal patient cohort. Further, all methods were able to classify the normal group against each of the diseased groups with statistical significance. IVPD ($p = <0.0001$ for N vs LVH, and $p = 0.0015$ for N vs DCM), $V_P$ ($p = <0.0001$ for N vs LVH, and $p = 0.0011$ for N vs DCM), $V_S$ ($p = <0.0001$ for N vs LVH, and $p = 0.0002$ for N vs DCM), Peak-$V_W$ ($p = <0.0001$ for N vs LVH, and $p <$ 0.0001 for N vs DCM). However, Peak-$V_W$ was able to more clearly delineate the diseased groups from the normal group. Furthermore, it was the only metric that better classified between DCM v LVH (p= 0.2171, Figure 5).

In addition, improved clinical utility of Peak-$V_W$ over IVPD, $V_P$, and $V_S$ was further supported by the increased AUC (Figure 6). In Figure 6(a), both remodeling geometries are treated as diseased and compared with the normal filling cohort. The Peak-$V_W$ (AUC=0.994) shows the highest AUC, followed by the filling strength $V_S$ (AUC=0.886), $V_P$ (AUC=0.864), and IVPD (AUC=0.851). The AUCs for all methods are statistically significant (p<0.0001). More importantly, the second ROC curve in Figure 6(b) differentiates the dilated ventricles from the hypertrophied ones. For this case as well, the highest AUC is obtained by the wavelet-based Peak-$V_w$ (AUC=0.864) and it is the most statistically significant prediction (p=0.0004) as compared to $V_S$ (AUC=0.655), $V_P$ (AUC=0.685) and IVPD (AUC=0.697).

## Discussion

We developed a new method based on wavelet analysis to calculate the LV filling wave propagation velocity. This new parameter, termed Peak-$V_W$, identifies the most energetic and dominant filling wave during diastolic flow and directly measures its propagation speed. Physically it captures the combined effect of the strength and energy of the LV suction as well as the effect of the shape of the LV. The shape of the LV, for example a dilated versus a hypertrophic LV, would exhibit different filling wave propagation velocities even under the same IVPD, since the distribution of momentum for the filling flow is shape-dependent. Although Peak-$V_W$ is conceptually parallel to the traditional $V_P$, we found that, it is able to better capture the physics, and thus may provide a more accurate and clinically useful evaluation of LV diastolic. Moreover, this method is objective, free of heuristics, and as a result has significantly lower inter- and intra-operator variability.

$V_P$ is conventionally measured as the slope of an iso-velocity contour of the blood velocity map, set at the 50% of the maximum inflow velocity from the mitral plane to a distance of 4 cm into the LV [1]. Done in this manner, $V_P$ provides a useful measure of LV relaxation. Furthermore, the ratio of the peak mitral valve inflow velocity (E) to $V_P$ (E /$V_P$) can be used to estimate the LV filling pressure. This is based on the assumption that the filling wave has a constant $V_P$ throughout early diastole. However, the iso-velocity contour is curvilinear indicating that the $V_P$ is not constant during early filling [12]. Moreover, the $V_P$ depends on the choice of aliasing boundary, corresponding to different iso-velocity contours [14,15], suggesting that early diastolic LV filling is not a bulk wave moving with a constant velocity but instead is comprised of a range of spatiotemporally varying propagation velocities. Hence, there are several potential limitations to the use of $V_P$ as a measure of LV diastolic function. In addition, $V_P$ is subject to variability based on how the isovelocity contour is determined (1,17,20). Finally, $V_P$ has been found to be normal in patients with hypertrophic cardiomyopathy (HCM) who have diastolic dysfunction apparent by other methods (21).

We previously attempted to address these limitations by describing $V_P$ by two slopes, corresponding to a rapid initial and a slower terminal $V_P$, which provided a more accurate representation of early filling. [13]. As a result, the product of the initial $V_P$ and the distance it travelled into the LV ($V_S$) provided a small improvement in assessing diastolic function.

To better account for these potential limitations to the conventional method of determining $V_P$, we developed a more physiologically robust approach. We used a continuous wavelet transform with varying time-step sizes for each disease stage to analyze the CMM LV filling wave without making other assumptions concerning the bulk transfer of blood into the LV. This enables the calculation of the propagation velocity for the most energetic wave $V_W$. Our new measure when calculated at the correct time step provided better recognition of diastolic function than the conventional measurement of $V_P$, or $V_S$, as well as IVPD. This improvement was apparent with both cohorts evaluated herein, namely in patients with diastolic dysfunction (Group A), and LV dilatation and LVH (Group B).

In Figure 7 we plot the peak-$V_W$ calculated using a spatial cross-correlation (without using wavelets) for a series of time steps (dt). The Normals and LVH peak VW variation shows a rise-decay trend while the Restrictives and DCM patients show a plateau trend (Normals, dt=0.012 Restrictives and LVH, dt=0.06; DCM, dt=0.04). The presence of spurious noise in the CMM scans causes a false peak $V_W$ in the diseased stages at a lower time step but they plateau to a constant peak $V_W$ for a higher time step. The physical significance of the "peak" and "plateau" trends is not fully understood yet. Ongoing work involves automating the algorithm to detect the correct time step for each scan before choosing the peak-$V_W$.

$V_P$ and IVPD have been developed as indices of LV suction, however their limitation under conditions of pseudo-normalization caused by elevated left atrial pressure has been documented [25]. This is also supported by our results, in which, $V_P$ and IVPD are higher in P than in IR. The newer index, $V_S$, also could not overcome this limitation in the present analysis. In contrast, Peak-$V_W$ measurements were significantly reduced in P compared to IR and N, indicating that this new approach to calculating propagation velocity could discriminate P from N, and that Peak-$V_W$ would be a robust marker of LV suction.

Moreover, Peak-$V_W$ could also differentiate LVH and DCM. In a hypertrophied heart, the suction is considered to be reduced due to impaired longitudinal LV function[26]. On the other hand, in a dilated heart, LV suction decreases owing to the flow disturbance in the enlarged chamber as well as the reduced recoil caused by systolic dysfunction [27]. These differences in the upstream of reduced LV suction could have resulted in the differences in Peak-$V_W$ between LVH and DCM. Although the discrimination between LVH from DCM using Peak-$V_W$ may not have an impact on the clinical practice, it could provide new insights for the understanding of complicated pathophysiology in LV diastolic function.

Overall, this study suggests that Peak-$V_W$ offers the clinical potential of determining progressively worsened diastolic function with minimal user input. This clinical ability may also be more useful when combination of traditional measures like mitral inflow velocities, tissue Doppler velocities, relaxation time and ejection fraction (E/A, E/E′, τ, EF ) are indeterminate due to arrhythmias [25,28,29] or regional wall motion variation (i.e. from myocardial infarction) [30] and when there are segmental differences in tissue Doppler velocities [31]. Emerging techniques of speckle tracking echocardiography measuring global strain and strain rate (SR) of the left ventricle provide incremental improvement in diagnostic capability of patients with normal ejection fraction but suffering from myocardial infarction [32,33]. In presence of diastolic dysfunction, circumferential SR was influenced by changes in IVPD but longitudinal SR remained unaffected by IVPD variations with an abnormal temporal non-uniformity causing delay of longitudinal relative to circumferential expansion [34,35].  Left atrium strain has recently shown a superior diagnostic capability to differentiate between all stages of diastolic dysfunction [36-38]. But all strain measurements have an inherent variability where the manual choice of segments on the chamber at which the strain-time history is extracted may vary from person to person. However, a combination of peak-Vw and LA strain as diagnostic metrics in future would encompass the cardiac tissue-flow physics completely and eliminate the need for any heuristic measures.

There are several potential limitations to our study. First, there is no ground-truth of diastolic function to use for validation of a new method. Second, we measured Peak-$V_W$ from a single beat, which means it does not account for any beat-to-beat variation. Third, the peak-Vw calculation is not completely agnostic yet as the correct time-step must be chosen for each scan before applying the continuous wavelet transform. Finally, we have assessed its clinical utility in a small, single center cohort. Although our new measurement may appear to involve complex calculations, this was accomplished with a semi-automated algorithm. Thus, it is possible that this algorithm be implemented online and be publicly available, offering the potential for diagnosing subclinical diastolic dysfunction with increased sensitivity.

## Acknowledgements

This material was originally supported under a National Institutes of Health R21 Grant No. HL106276-01A1. The authors have no conflicts of interest to report.

# Figures

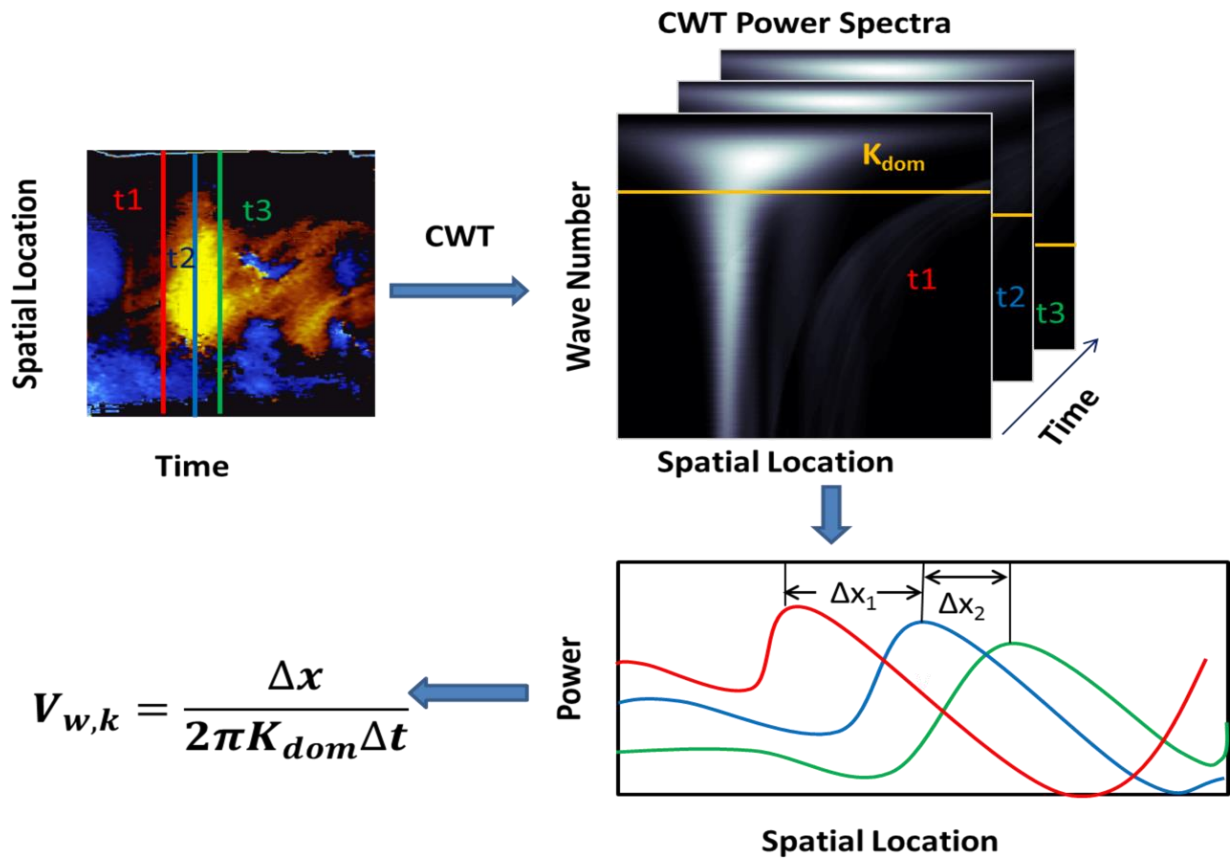

Figure 1:  On the top left three spatially varying signals corresponding to times t1, t2 and t3 are shown on a CMM scan. The CWT transform for each time signal provides the power spectra shown in the top right figure, from which the dominant wavenumber  is determined. The bottom right figure shows a schematic of a representative spatial variation of power for each time signal at its dominant wavenumber. The spatial lag ( $\Delta$ x) shown in the figure, the known time difference ( $\Delta$ t) and the dominant wavenumber ($K_{dom}$) combine to give the peak propagation velocity, $V_{w,k.}$

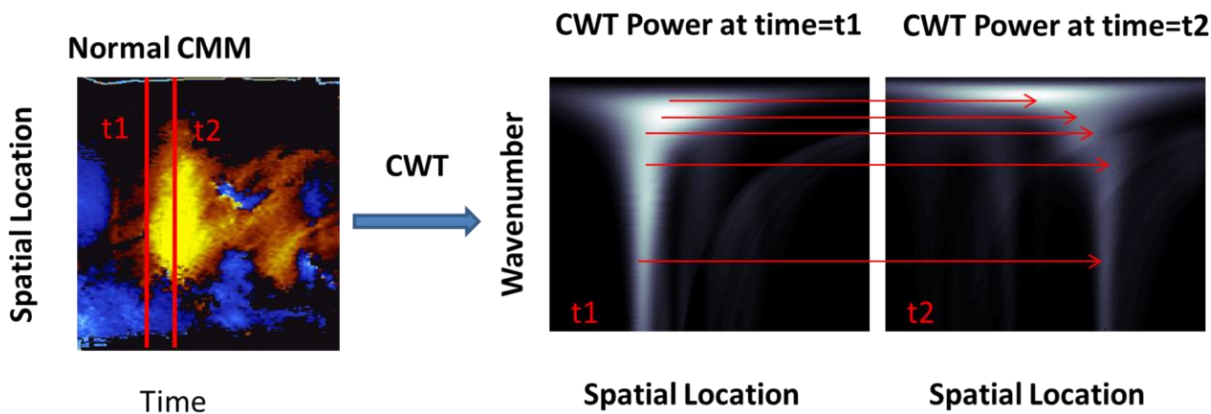

Figure 2: Representative wavelet analysis. The first column shows the CMM echocardiogram for a normal filling representative patient with *t1* and *t2* indicated by the vertical red lines. The next two columns show the CWT power spectra at *t1* and at *t2*. The red lines show the motion of the peaks at different wavenumbers. The CWT is applied to each temporal velocity signal on the CMM echocardiogram resulting in a CWT power spectrum localized in space, time and wavenumber. The spatial lag of the peak on the CWT power spectra is calculated between two time steps, for each wavenumber (horizontal location). The spatial lag and the time between the two time signals are used to calculate a propagation velocity for each wavenumber. As is evident from the CWT images, higher wavenumbers have higher displacements than lower wave numbers. The range of dominant wavenumbers is specific to each patient. The minimum wave number considered is the inverse of the physical length of the LV. The maximum wavenumber is also specific to each patient and is dependent on the most energetic wave components present in each signal.

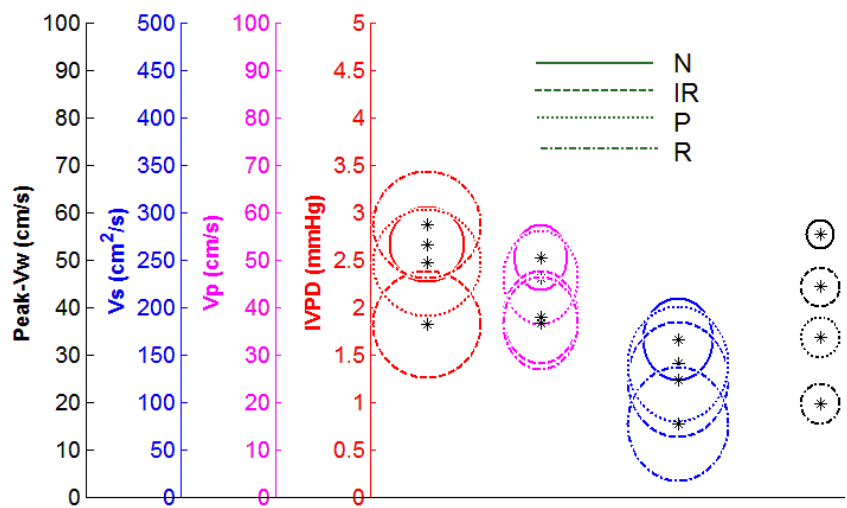

Figure 3: Color-M-Mode based parameters for each patient cohort for group A. The black star at the circle center denotes the median value of each parameter in each cohort. The radius of the circle denotes the deviation in each parameter. Color of the deviation circles denotes methodology and corresponds to axis colors.

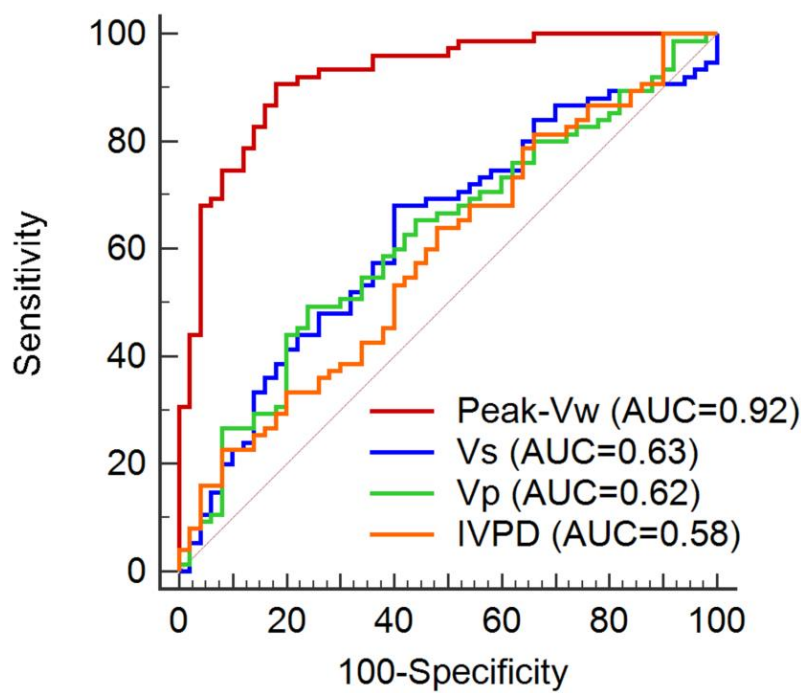

Figure 4: Receiver Operator Characteristic (ROC) curves for three conventional parameters and for the proposed wavelet parameter of Group A. The area under the curve (AUC) is noted in the legend for each parameter. The plot differentiates the normal filling cases from diseased filling ones. The maximum wave component peak-$V_w$ displays the highest AUC.

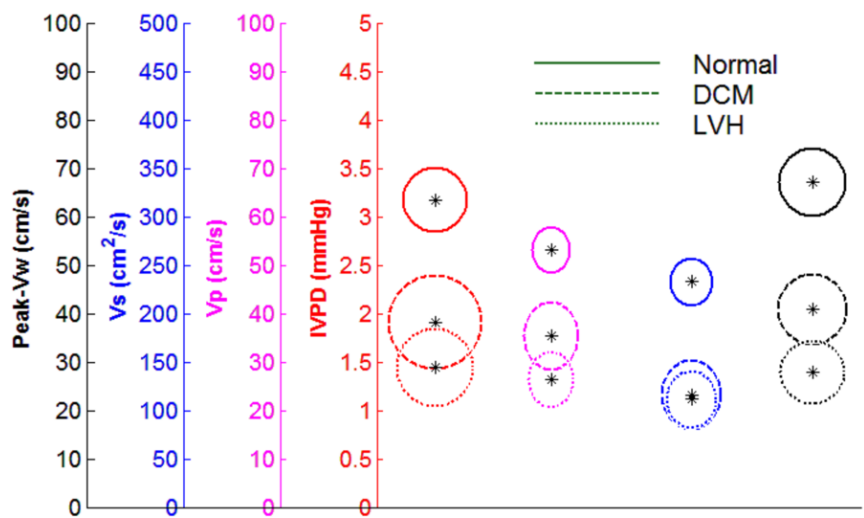

Figure 5: CMM based parameters for each patient cohort in group B. The black star at the circle center denotes the median value of each parameter in each cohort. The radius of the circle denotes the deviation in each parameter. Color of the deviation circles denotes methodology and corresponds to axis colors.

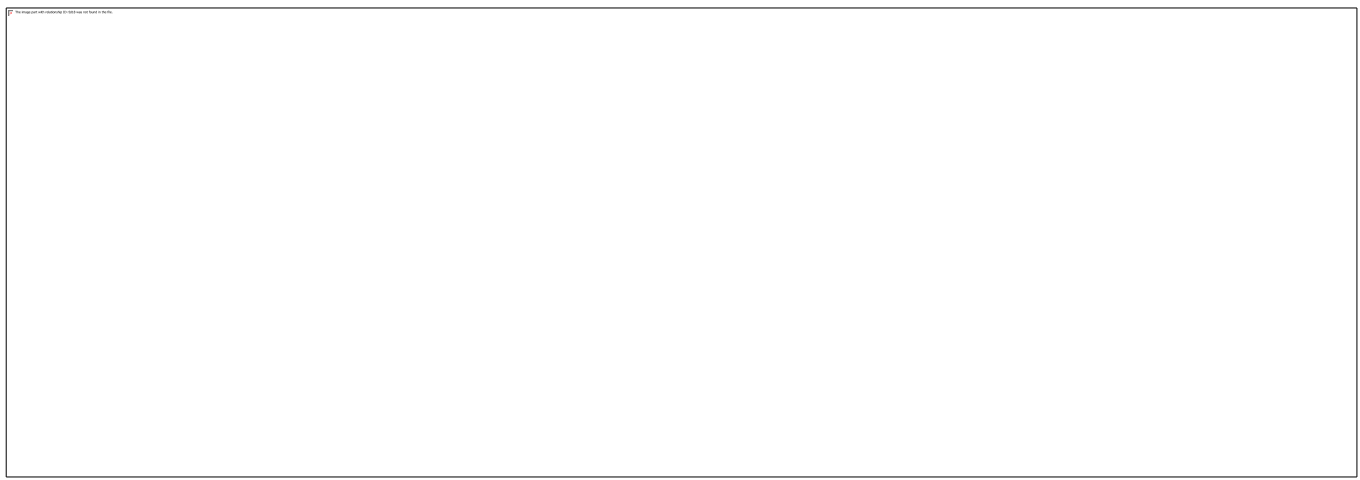

(a)                                                    (b)

Figure 6: Operator Characteristic (ROC) curves for all methods of the Group B. The area under the curve (AUC) is noted in the legend for each parameter. (a) The classification is based on normal ventricle geometry versus remodeled ones. (b) The classification is based on LVH ventricle geometry versus DCM ones for the plot on the right.

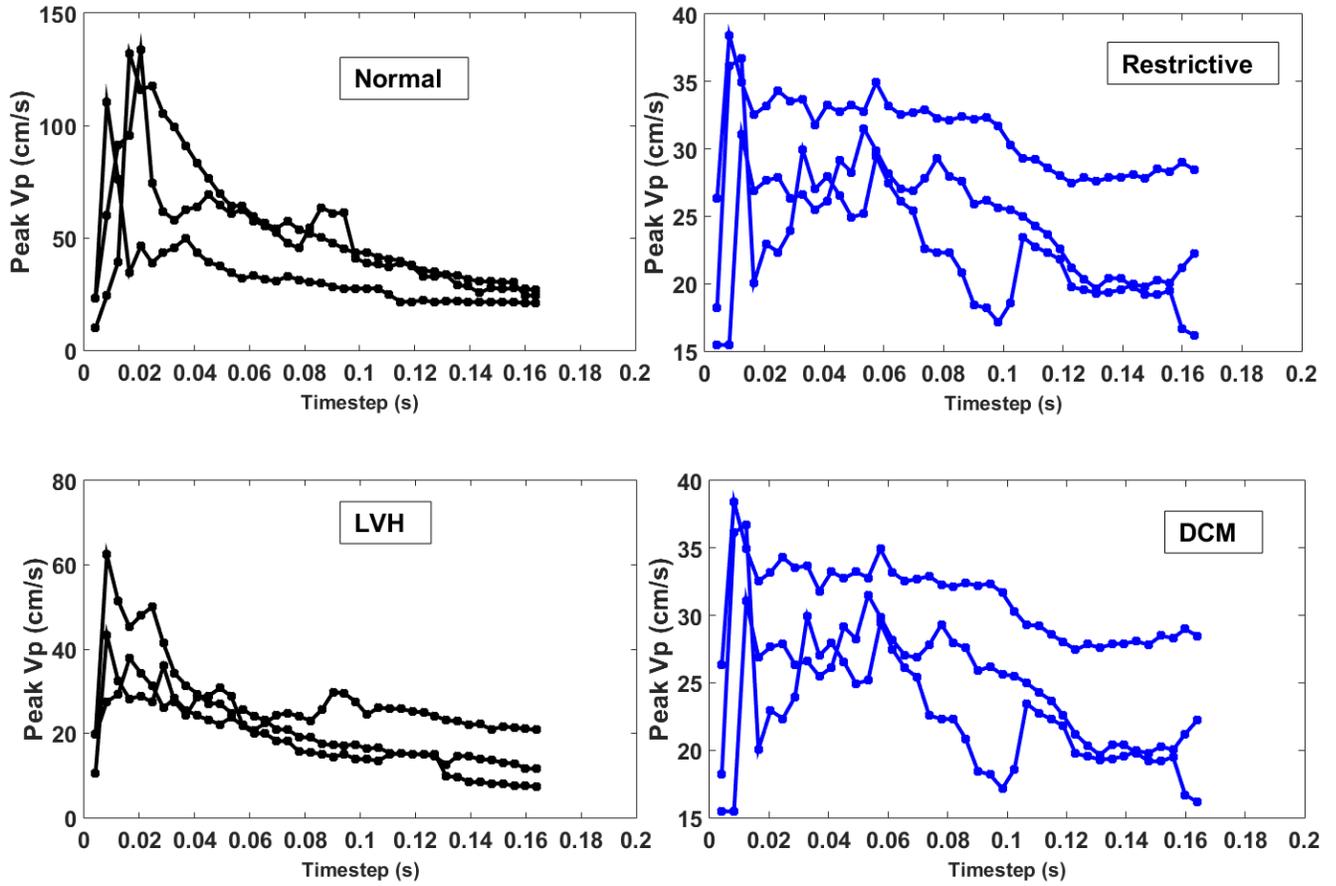

Figure 7: Variation of peak Vp with changing time step size shown for 3 sample patients in each of the Normal, Restrictive, LVH and DCM cohorts.

# Tables

Table 1: Group A patient characteristics

| Diastolic Dysfunction Stage | Number of Patients | Age (years) | E/A | E/E' | Ejection Fraction |
|---|---|---|---|---|---|
| N (Normal) | 50 | 38.6±15.1 | 1.8±.4 | 6.8±1.9 | 0.6±0.1 |
| IR (Stage 1 - Impaired relaxation) | 25 | 68.3±9.6 | 0.8±0.1 | 13.0±4.2 | 0.6±0.1 |
| P (Stage 2- Pseudonormal) | 25 | 66.2±12.9 | 1.6±0.3 | 16.4±5.3 | 0.4±0.2 |
| R (Stage 3- Restrictive) | 25 | 59.4±18.5 | 2.9±1.0 | 18.8±7.4 | 0.3±0.1 |

*Patients are classified based on clinically diagnosed diastolic function stage. Values represent mean±SD (standard deviation).
† E/A= E–wave to A-wave transmitral velocity ratio;
‡E/E'= transmitral Doppler E-wave velocity to mitral annular tissue Doppler E-wave velocity ratio.

Table 2: Group B patient characteristics

| Subgroup | Number of patients | Age (years) | | Ejection Fraction | | Geometry Classifier |
|---|---|---|---|---|---|---|
| | | Median | IQR | Median | IQR | |
| N(Normal) | 32 | 30 | 24 | 60 | 8 | Normal |
| D (DCM) | 15 | 58 | 15 | 20 | 10 | SI < 1.60 |
| H (LVH ) | 22 | 60 | 9 | 16 | 7 | LVMI > 175 |

* Values reported are median and interquartile range, IQR (75% value - 25% value)
† LVMI-Left Ventricular Mass Index
‡ SI-Sphericity Index

Table 3: Results of intra-observer variability and inter-observer variability for each parameter. Peak-$V_W$ reports the smallest variability when compared to conventional parameters.

| Coefficient of variation (%) | $V_S$ | $V_P$ | Peak-$V_W$ | IVPD |
|---|---|---|---|---|
| Median Intra-observer | 12 | 10.5 | 3.7 | 5.8 |
| Median Inter-observer | 19.6 | 17.2 | 10.1 | 14.5 |

Table 4: Summary of results for Group A.

| | IVPD (mmHg) | | V$_P$ (cm/s) | | V$_S$ (cm$^2$/s) | | Peak-V$_W$ (cm/s) | |
|---|---|---|---|---|---|---|---|---|
| | Median | IQR | Median | IQR | Median | IQR | Median | IQR |
| N | 2.7 | 1.9 | 50.7 | 26.9 | 167.1 | 148.7 | 55.4 | 12.4 |
| IR | 1.8 | 1.5 | 37.9 | 39.9 | 123.73 | 163.7 | 44.3 | 10.2 |
| P | 2.5 | 1.9 | 46.2 | 31.4 | 140.3 | 138.8 | 33.7 | 11.7 |
| R | 2.9 | 1.9 | 36.7 | 17.9 | 76.7 | 87.4 | 19.6 | 19.0 |

Table 5: p-values from Tukey-Kramer HSD test for Group A. The shaded values signify statistical significance. The wavelet based peak propagation velocity has highest statistical significance.

| Parameter | Filling Pairs | | p-value |
|---|---|---|---|
| **IVPD (mmHg)** | N | IR | 0.0024* |
| | N | P | 0.8887 |
| | N | R | 0.9975 |
| | IR | P | 0.0628 |
| | IR | R | 0.0185* |
| | P | R | 0.9681 |
| **V$_P$ (cm/s)** | N | IR | 0.2111 |
| | N | P | 0.9558 |
| | N | R | 0.1008 |
| | IR | P | 0.5975 |
| | IR | R | 0.9896 |
| | P | R | 0.4064 |
| **V$_S$ (cm$^2$/s)** | N | IR | 0.9986 |
| | N | P | 0.7375 |
| | N | R | 0.0644 |
| | IR | P | 0.7387 |
| | IR | R | 0.1816 |
| | P | R | 0.0146* |
| **Peak-V$_W$ (cm/s)** | N | IR | <0.0001* |
| | N | P | <0.0001* |
| | N | R | <0.0001* |
| | IR | P | 0.006* |
| | IR | R | <0.0001* |
| | P | R | 0.0015* |

Table 6: Summary of results for Group B.

| | IVPD (mmHg) | | $V_P$ (cm/s) | | $V_S$ (cm$^2$/s) | | Peak-$V_W$ (cm/s) | |
|---|---|---|---|---|---|---|---|---|
| | Median | IQR | Median | IQR | Median | IQR | Median | IQR |
| N | 3.2 | 2.0 | 53.1 | 24.4 | 232.8 | 153.4 | 67.2 | 23.2 |
| LVH | 1.4 | 1.3 | 26.4 | 13.8 | 111.4 | 89.2 | 27.8 | 14.1 |
| DCM | 1.9 | 0.9 | 35.3 | 23.8 | 116 | 132.6 | 40.8 | 5.9 |